\documentclass[pra,twocolumn,aps,groupedaddress,longbibliography,nofootinbib]{revtex4-1}
\usepackage{hyperref}
\usepackage{graphicx}
\usepackage{amsmath}
\usepackage{amsfonts}
\usepackage{amssymb}
 \usepackage{mathtools}
\usepackage{epsfig}
\usepackage{subfigure}
\usepackage{mathtools}
\usepackage{braket}
\usepackage{float}
\usepackage[usenames,dvipsnames]{color}
\usepackage{setspace}
\usepackage{bm}
\usepackage{times}
\hypersetup{
      colorlinks=true,
      citecolor=blue,
      linkcolor=blue,
      urlcolor=blue}

\definecolor{orange(colorwheel)}{rgb}{1.0, 0.5, 0.0}
\definecolor{maroon(html/css)}{rgb}{0.5, 0.0, 0.0}
\definecolor{lightgray}{gray}{0.9}

\newcommand{\Ut}{\tilde{U}}
\newcommand{\br}{\mathbf{r}}
\newcommand{\brho}{{\boldsymbol{\rho}}}
\newcommand{\bk}{\mathbf{k}}
\newcommand{\MF}{\mathcal{F}}

\begin{document}
\title{Excitations and number fluctuations in an elongated dipolar Bose-Einstein condensate} 
\author{Sukla Pal, D. Baillie and P. B. Blakie}
\affiliation{Department of Physics, Centre for Quantum Science, and The Dodd-Walls
            	Centre for \\ Photonic and Quantum Technologies, University of
            	Otago, Dunedin 9016, New Zealand}
         \date{\today}
          
         \begin{abstract}
         	
         	We study the properties of a magnetic dipolar Bose-Einstein condensate (BEC) in an elongated (cigar shaped) confining potential in the beyond quasi-one-dimensional (quasi-1D) regime. In this system the dipole-dipole interactions (DDIs) develop a momentum-dependence related to the transverse confinement and the polarization direction of the dipoles.  
	This leads to density fluctuations being enhanced or suppressed at a length scale related to the transverse confinement length, with local atom number measurements being a practical method to observe these effects in experiments. 
	 We  use meanfield theory to describe the ground state, excitations and the local number fluctuations.  Quantitative predictions are presented based on full numerical solutions and a simplified variational approach that we develop. In addition to the well-known roton excitation, occurring when the dipoles are polarized along a tightly confined direction, we find an ``anti-roton" effect for the case of dipoles polarized along the long axis: a nearly non-interacting ground state that experiences strongly repulsive interactions with excitations of sufficiently short wavelength.
       	
                  \end{abstract}
         \maketitle

\section{Introduction}
         {Ultracold atomic gases are  superb systems for  exploring manybody physics with well-characterized and controllable microscopic parameters. Recently dipolar BECs have been used to reveal a roton-like collective mode \cite{Chomaz2018a}. 
      This excitation was prepared by tuning the short ranged contact interactions when the  condensate was confined in an elongated geometry with the atomic dipoles polarized along one of the tightly confined directions (e.g.~see $\alpha=\pi/2$ case in Fig.~\ref{schematic}). Evidence for the roton was provided by the observation of unstable dynamics \cite{Chomaz2018a} and using Bragg spectroscopy to probe the excitations \cite{Petter2019a}. If the roton is tuned to zero energy the elongated dipolar condensate can transition into a supersolid state of matter (i.e.~spontaneously modulated along $z$), as observed recently in three labs \cite{Tanzi2019a,Bottcher2019a,Chomaz2019a}. In order to observe these phenomena the experiments need to be in the regime where the interaction energies are much larger than the zero point energy of the transverse confinement (i.e.~far outside the quasi-1D regime).

 Fluctuation measurements in quantum systems can be used to reveal the interplay of quantum statistics and interactions. In quantum gases it is of interest to make local measurements of fluctuations \cite{Giorgini1998a,Belzig2007a,Astrakharchik2007a,Abad2013a}, e.g.~using optical imaging to measure atom number statistics in a small region of interest, as have been demonstrated in a number of experiments  \cite{Jacqmin2011a,Armijo2011a,Armijo2012a,Hung2011,Hung2011a,Hung2013a,Blumkin2013a,Schley2013a}. 
There has also been theoretical interest in the local fluctuations of quasi-two-dimensional (or pancake-shaped) dipolar BECs \cite{Klawunn2011a,Bisset2013a,Bisset2013c,Baillie2014a}. A key prediction being that a roton excitation causes the condensate to exhibit large fluctuations for measurement cells of size comparable to the roton wavelength.   

         \begin{figure}[tbh]
         	\centering
         	\includegraphics[width=3.2in]{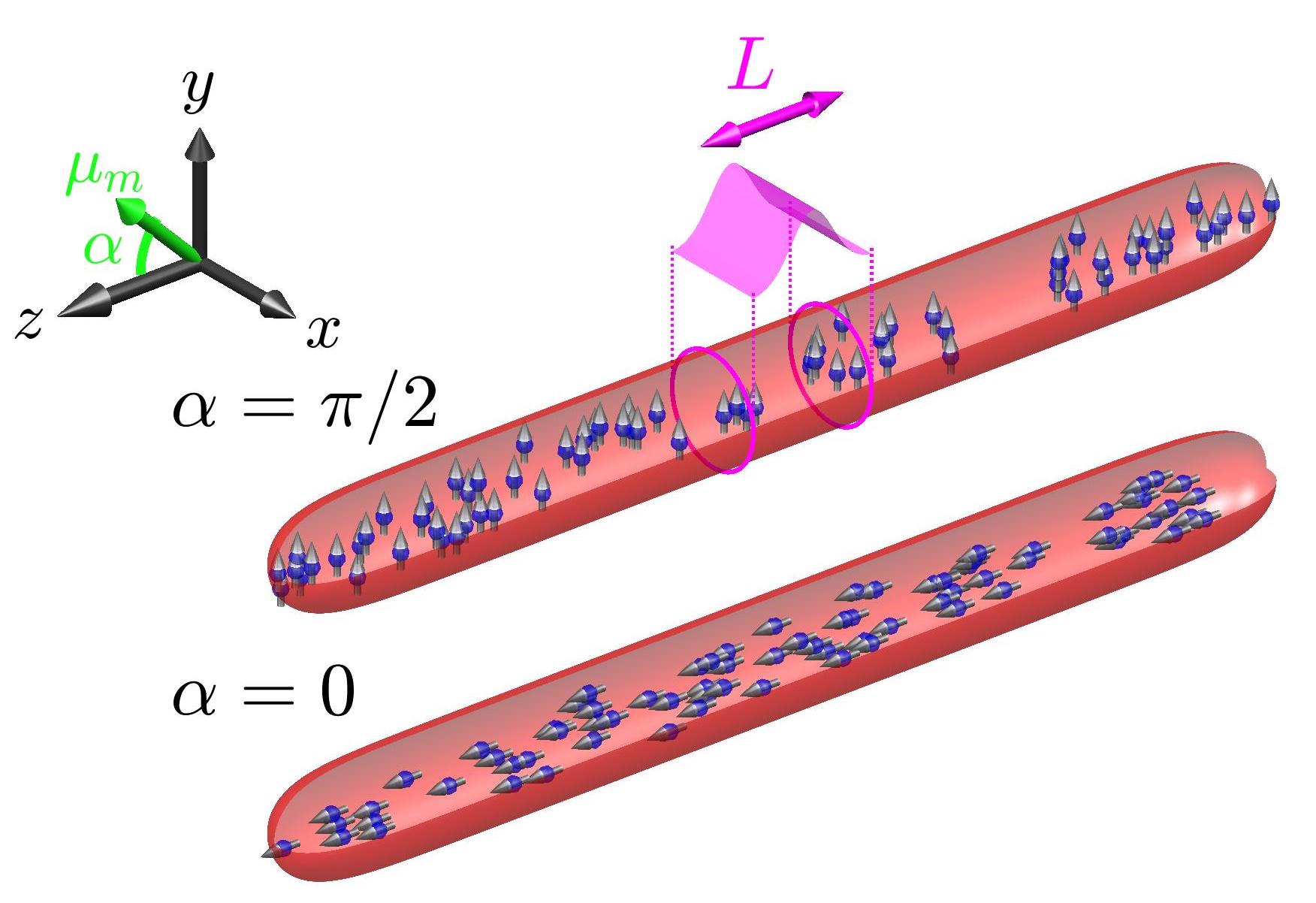}    	
         	\caption{Schematic diagram: a one dimensional Gaussian cell of length $L$ (shown by magenta surface profile) along the axial direction ($z$-axis) is used to measure atom number in an elongated dipolar BEC.  The dipole magnetic moment ${\boldsymbol \mu_m}$ makes angle $\alpha$ with the axial direction, and we show cases where the dipoles are polarized along ($\alpha = 0$) and transverse ($\alpha = \pi/2$) to the long axis.}
         	\label{schematic}
         \end{figure}
       
In this paper we develop formalism for describing the ground state and excitations of an elongated dipolar condensate in the beyond-quasi-1D regime relevant to current experiments. To simplify our treatment we neglect trapping along the weakly confined direction (along the $z$-axis in Fig.~\ref{schematic}), and take the system to be uniform. We also develop a variational approach (justified by comparison to full numerical calculations) for computing the system properties.

 The variational approach also affords us insight into the momentum dependence (along the $z$-axis) of interactions and its effect on the excitations. This momentum dependence arises from the DDIs under the transverse confinement, thus is sensitive to the dipole orientation and introduces the transverse confinement as a new length scale for the system.
  For the case of dipoles polarized along the tightly confined direction the interactions are strongly repulsive at long wavelengths (i.e.~a high speed of sound) and weakly repulsive or attractive at wavelengths comparable to, or shorter than, the confinement length. Indeed, for sufficiently strong DDIs this latter effect leads to the formation of the roton excitation, i.e.~the attractive character of the interactions causes a local minima in the excitation dispersion relation.  Interestingly the contrasting case of dipoles polarized along the long axis has received less attention. Here the interaction is weakly repulsive at long wavelengths (i.e.~a low speed of sound), but becomes strongly repulsive at wavelengths shorter than the confinement length. Indeed, in this case the ground state experiences a weak interaction, yet a strong repulsive interaction occurs between the condensate and short wavelength excitations.  We refer to this as an ``anti-roton" effect, in that it causes a significant upward shift in the energy of higher-momentum excitations. Interestingly, this effect also causes the dynamic structure factor (characterising the dynamics of the density fluctuations) to spread its weight over many bands.

 Finally we apply our theory to study the local number fluctuations, determined by measuring the number of atoms in a small Gaussian cell of finite length $L$, as indicated in Fig.~\ref{schematic}. Such measurements can be  made  in experiments using finite resolution absorption imaging, with the length scale $L$ being adjustable by merging the results of camera pixels (e.g.~see Ref.~\cite{Armijo2012a}). We show that these measurements can be used to reveal the key features of the excitations and fluctuations of an elongated dipolar condensate. Our findings should serve as motivation for experiments to make the first fluctuation measurements of a dipolar BEC.

         The outline of the paper is as follows: We begin in Sec.~\ref{formaism} with a general discussion of fluctuation measurements in ultra-cold gases. We also describe the meanfield formalism for the ground state and excitations of the elongated dipolar system, and develop a variational approximation.   Following that in Sec.~\ref{sec:cases} we consider the ground state, excitations and structure factors for several cases that illustrate the system properties, including the anti-roton effect and the development of a roton excitation.  In Sec.~\ref{results} we present results for the atom number fluctuations and make comparisons to the variational approach. Finally, we conclude in Sec.~\ref{conclusions}.

\section{Formalism} \label{formaism} 
\subsection{Basic theory of local number fluctuations}
        
Previous experiments with an elongated Rb BEC \cite{Jacqmin2011a,Armijo2011a,Armijo2012a}  have measured number fluctuations using \textit{in situ} absorption imaging in a manner similar to what we consider here. The finite imaging resolution  (c.f.~the Gaussian surface representing the imaging system point spread function in Fig.~\ref{schematic}) means that these measurements effectively count the number of atoms within a cell. Here we take the cell to be described by the weight function  $\Lambda(z) = \sqrt{2}e^{-z^2/L^2}$, representing an imaging system has a resolution length of $L$ along  $z$, and the transverse directions are considered completely unresolved (cf.~\cite{Jacqmin2011a,Armijo2011a,Armijo2012a}). In practice the minimum value of $L$   ($\sim1\mu$m$)$ is set by the details of the imaging system, but larger values can be trivially arranged by combining the measurements from neighboring cells. The observable of interest is the number of atoms in the cell, which for the case of a cell centred at the origin is given by
\begin{align}
\hat{N}_L=\int dz\,\Lambda(z)\hat{n}(z),
\end{align}
where $\hat{n}(z)=\int d\bm{\rho}\,\hat{\psi}^\dagger(\mathbf{x})\hat{\psi}(\mathbf{x})$ is the linear density operator, with   $\hat{\psi}$ being the field operator for the system and $\bm{\rho}\equiv(x,y)$ representing the transverse coordinates. Here we will consider the case of a dipolar BEC that is translationally invariant along $z$ such that the linear density is uniform, i.e.~$n=\langle\hat{n}(z)\rangle$. 
The fluctuations in the measurement of the number of atoms in the cell is given by $\Delta N_{L}^2 =  \langle (\hat{N}_{L} - N_{L})^2\rangle$, where $N_{L}=\langle \hat{N}_L\rangle=\sqrt{2\pi}nL$ is the mean atom number.
For the translationally invariant system this can be evaluated as
\begin{align}
\label{var}
         \Delta N_{L}^2 =  n\int \frac{dk_z}{2\pi} \tilde{\tau}_{\Lambda}(k_z) S(k_z), 
\end{align}
where  $\tilde{\tau}_{\Lambda}(k_z)= 2\pi L^2 e^{-k_z^2L^2/2}$ is the Fourier transform of cell geometry function $\tau_{\Lambda}(\Delta) = \int dz\int dz' \Lambda(z)\Lambda(z')\delta(z-z'-\Delta) = \sqrt{2\pi}L e^{-\Delta^2/2L^2} $. We have also introduced the static structure factor
\begin{align}
S(k_z)=\frac{1}{n}\int dz\langle\delta\hat{n}(z)\,\delta \hat{n}(0)\rangle e^{-ik_zz},
\label{Skz}
\end{align}
as the Fourier transform of the density correlation function, where $\delta \hat{n}(z)=\hat{n}(z)-n$ is the density fluctuation operator. When the system exhibits density correlations at a length scale comparable to the cell size, we expect these to be revealed in the measured number fluctuations. Indeed, a focus here will be the influence of density correlations arising from the interplay of the DDI and transverse confinement.

For a cell that is much longer than any of the relevant correlation lengths\footnote{This includes the density healing length, thermal coherence length and any additional length scales introduced by the interactions.} of the system, the number fluctuations approach a thermodynamic limit  that is universal for compressible superfluids \cite{Astrakharchik2007a} (also see \cite{PitaevskiiStringariBook,Klawunn2011a})
        \begin{equation}\label{thlimit}
         \Delta N_{L}^2 = N_L\frac{k_BT}{mc^2},
         \end{equation}
         where  $c$ is the speed of sound, $T$ is the temperature and $m$ is the atomic mass.
          
          On the other hand, for a cell that is much shorter than all relevant correlation lengths, the particles can be treated as independent, which gives rise to Poissonian statistics for the fluctuations, i.e.~$\Delta N_L^2 = N_L$. Equivalently [from Eq.~(\ref{var})] this arises from the static structure factor having the (uncorrelated) high-$k_z$ limit $S\to1$. So far it has not been practical for experiments to make measurements in this limit using optical imagining, as this would typically require $L\ll 1\mu$m.

%
\subsection{Meanfield Formalism}

     Here we outline the meanfield description appropriate to a dipolar BEC under transverse harmonic confinement. We introduce the relevant Gross-Pitaevskii and Bogoliubov de-Gennes formalism, and briefly describe the numerical methods we employ to solve the associated equations.
     
\subsubsection{Gross-Pitaevskii theory}
The condensate field $\psi(\mathbf{x})=\langle\hat{\psi}(\mathbf{x})\rangle$ satisfies the non-local Gross-Pitaevskii equation (GPE) $\mathcal{L}_{GP}\psi = \mu \psi$, where
     \begin{align}\label{gp}
     \mathcal{L}_{GP} \equiv -\frac{\hbar^2\nabla^2}{2m}\! +\!V(\bm{\rho}) \! + \! \int\! d{\bf x}'U({\bf x}-{\bf x'})|\psi({\bf x'})|^2, 
     \end{align}
$V(\bm{\rho})=\frac{1}{2}m(\omega_x^2x^2+\omega_y^2y^2)$ is the transverse harmonic confinement, with angular frequencies $\omega_x$ and $\omega_y$ along the two transverse directions, and $\mu$ is the chemical potential.
  The two-body interactions are described by the potential 
  \begin{align}
  U({\bf r}) = g_s \delta({\bf r}) + U_{dd}({\bf r}),\label{Urfull}
  \end{align}
which contains the short ranged contact interaction with coupling constant $g_s = 4\pi\hbar^2 a_s/m$, where $a_s$ is the $s$-wave scattering length.  The long ranged DDI is described by 
   \begin{eqnarray}\label{ddi}
     U_{dd}({\bf r}) =  \frac{3g_{dd}}{4\pi r^3} [1 - 3(\hat{\mathbf{r}}\cdot\hat{\bm{\mu}}_m)^2],
     \end{eqnarray} 
     where the coupling constant $g_{dd} = 4\pi\hbar^2 a_{dd}/m$ introduces the dipole length $a_{dd}$ that relates to the magnetic moment of the atoms as $a_{dd}= \mu_0\mu_m^2 m /12 \pi \hbar^2$. The form of the DDI depends on the dipole polarization direction. Here we take the dipole moment $\bm{\mu}_m$ to lie in the $yz$-plane at an angle of $\alpha$ to the $z$-axis, i.e.~$\bm{\mu}_m=\mu_m(\cos\alpha\hat{\mathbf{z}}+\sin\alpha\hat{\mathbf{y}})$,  (see Fig.~\ref{schematic}). However we present results here only for the two extreme cases of $\alpha=0$ and $\pi/2$.

 We restrict our attention to ground states that are uniform along $z$ (i.e.~we do not consider the possibility of ground states that break the translational symmetry, such as supersolids). Thus the ground state solution takes the form $\psi({\bf x}) =  \sqrt{n} \chi({\boldsymbol \rho})$, where $\chi$ is the unit normalized transverse mode and $n$ is the specified linear density.     In this case the GPE simplifies to 
  \begin{align}
  \mathcal{L}_{GP}^\perp\chi=\mu\chi,\label{GPE3D}
  \end{align}
   where
  \begin{align}
      \mathcal{L}^{\perp}_{GP} \equiv -\frac{\hbar^2\nabla_{\bm{\rho}}^2}{2m} + V({\boldsymbol \rho}) +  n\MF_\brho^{-1}\{\Ut(\bk_\brho)\MF_\brho\{[\chi(\brho)]^2\}\} ,\label{LGPperp}
  \end{align}
  with $\MF_\brho$ the Fourier transform in $\brho$, and $\Ut(\bk) = \int d\br\,e^{-i\bk\cdot\br} U(\br)$.

\subsubsection{Bogoliubov de-Gennes formalism}
     We also want to consider the elementary excitations of the condensate within the framework of  Bogoliubov theory \cite{Ronen2006a, PitaevskiiStringariBook}. This allows us to express the quantum field operator as
     \begin{align}
     \hat{\psi} =\sqrt{n}\chi +\sum_{k_z,j}\left[u_{k_zj}e^{-i{k_z}z}\hat{b}_{{k_zj}}-v_{k_zj}^*e^{i{k_z}z}\hat{b}_{{k_z}j}^{\dagger}\right],
     \end{align}
     where $\{u_{k_zj}(\bm{\rho}),v_{k_zj}(\bm{\rho})\}$ are the quasiparticle modes with respective eigenvalues $\{E_{k_zj}\}$ and bosonic mode operators $\{\hat{b}_{{k_zj}},\hat{b}_{{k_zj}}^\dagger\}$ satisfying the commutation relations $[\hat{b}_{{k_zj}},\hat{b}_{k_z'j'}^\dagger]=\delta_{k_zk_z'}\delta_{jj'}$.  The quantum number $k_z$ characterizes the wavevector of the quasiparticle along $z$, while the quantum number $j$ describes the transverse degrees of freedom.
     The quasiparticle properties are determined by solving the Bogoliubov de-Gennes equations  
     \begin{align}\label{bdgm}
      E_{k_zj}\begin{pmatrix} u_{k_zj}\\ v_{k_zj}  \end{pmatrix} = 
      \begin{pmatrix} \mathcal{L}_{k_z} -\mu& -X_{k_z}\\ X_{k_z} & - (\mathcal{L}_{k_z}-\mu) \end{pmatrix}
      \begin{pmatrix} u_{k_zj}\\ v_{k_zj}\end{pmatrix},
      \end{align} 
      where, $\mathcal{L}_{k_z} \equiv \mathcal{L}^{\perp}_{GP} + \epsilon_{k_z} +X_{k_z}$, $\epsilon_{k_z} =\hbar^2k_z^2/2m$,   
      and the exchange term is
      \begin{eqnarray}\label{x}
      X_{k_z} f(\brho)&\equiv& n\chi(\brho) \MF^{-1}_\brho\{\Ut(\bk) \MF_\brho\{\chi(\brho)f(\brho)\}\}.
      \end{eqnarray}
The speed of sound can then be identified from the slope of the lowest excitation band in the long wavelength limit, i.e.~as 
\begin{align}
c_\alpha=\left.\frac{1}{\hbar}\frac{\partial E_{k_z,j=0}}{\partial k_z}\right|_{k_z=0}.\label{cBdG}
\end{align}
We have explicitly labelled the speed of sound with the dipole angle $\alpha$ for future convenience.

From the solution of the GPE (\ref{GPE3D}) and BdG equations (\ref{bdgm}) we can calculate the dynamic and static structure factors of the system. While our primary interest is in the static structure factor it is useful to first introduce the dynamic structure factor, which describes the system response to a density coupled probe. For a dilute condensate  this can be evaluated as \cite{Zambelli2000a,Blakie2002a}
\begin{align}
 S(k_z,\omega)=\sum_j|\delta n_{k_zj}|^2[1+2f(E_{k_zj})]\delta(\omega-E_{k_zj}/\hbar),\label{DSF}
\end{align}
where 
\begin{align}
\delta n_{k_zj} = \int d{\boldsymbol \rho}\big[u_{k_zj}^{\ast}({\boldsymbol {\rho}}) - v_{k_zj}^{\ast }({\boldsymbol {\rho}})\big]\chi ({\boldsymbol \rho}),\label{deltank}
\end{align}
and $f(E)=(e^{E/k_BT}-1)^{-1}$ is the Bose factor. In cold-gas experiments  Bragg spectroscopy (e.g.~see \cite{Stenger1999a,StamperKurn1999a}) is sensitive to the zero temperature dynamic structure factor [i.e.~Eq.~(\ref{DSF}) with $f\to0$] and Petter \textit{et al.}~\cite{Petter2019a} have used Bragg spectroscopy on an elongated dipolar condensate to quantify the roton excitation spectrum.
 The static structure factor is obtained by integrating $S(k_z,\omega)$ over frequency, yielding 
\begin{align}
S({k_z})=\sum_{j}\left|\delta n_{k_zj}\right|^2[1+2f(E_{k_zj})].\label{SqBdG}
\end{align}

 \subsubsection{Numerical methods}   
 
We solve the GPE  (\ref{GPE3D})
by discretizing $\chi(\bm{\rho})$ on a two-dimensional numerical grid and using discrete Fourier transforms to apply the kinetic energy operator. We also use a Fourier transform to evaluate the interaction term [i.e.~convolution in Eq.~(\ref{LGPperp})] in $k$-space, but with a cutoff-interaction to improve accuracy (e.g.~see \cite{Ronen2006a,Lu2010a}). The GPE is solved using a gradient flow technique \cite{Bao2010a}.  The BdG equations are solved using large-scale eigensolvers.

For the $\alpha=0$ case with $\omega_x=\omega_y$ the problem is cylindrically symmetric, and $\chi$ reduces to being a function of $\rho=|\bm{\rho}|$. We can then solve for the ground state and the excitations on a one-dimension numerical grid using Bessel transformations \cite{Ronen2006a}.

\subsection{Variational theory}
     
  A simple variational theory for the elongated dipolar BEC has been developed in Ref.~\cite{Blakie2020a,Blakie2020b} for the case $\alpha=\pi/2$. Here we briefly review this theory and extend it to the $\alpha=0$ case.
  The basis of the variational theory is to approximate the transverse mode of the condensate     using  the Gaussian  $\chi^\text{var}(\bm{\rho})=\tfrac{1}{\sqrt{\pi}l} {e^{-(\eta x^2+y^2/\eta)/2l^2}}$, with variational parameters $l$ and $\eta$ describing the mean width and anisotropy, respectively.   
For our case (uniform density $n$ along $z$) this formalism reduces to a simple energy functional  
\begin{align}
\mathcal{E}_u(l,\eta)  =\mathcal{E}_\perp+ \tfrac{1}{2}n \tilde{U}_\alpha(0),\label{Eq:Eu}
 \end{align}
 where
 \begin{align}
 \mathcal{E}_\perp = \frac{\hbar^2}{4ml^2}\left(\eta+\frac{1}{\eta}\right)+\frac{ml^2}{4}\left(\frac{\omega_x^2}{\eta}+\omega_y^2\eta\right),
 \end{align}
 is the single-particle energy associated with the transverse degrees of freedom, 
 and
 \begin{align}
 \tilde{U}_\alpha(0)=\frac {1}{2\pi l^2}\left(g_s+g_{dd}\frac{2-\eta-3\cos^2\alpha}{1+\eta}\right),\label{U0}
 \end{align}
 is the effective interaction\footnote{Our notation here preempts that this is the $k_z=0$ limit of the $k_z$-space interaction we introduce later. } obtained by integrating out $\chi^\text{var}$. The parameters $\{l,\eta\}$ are determined by minimising Eq.~(\ref{Eq:Eu}).

 \begin{figure}[tbh]
      	\centering
      	\includegraphics[width=3.3in]{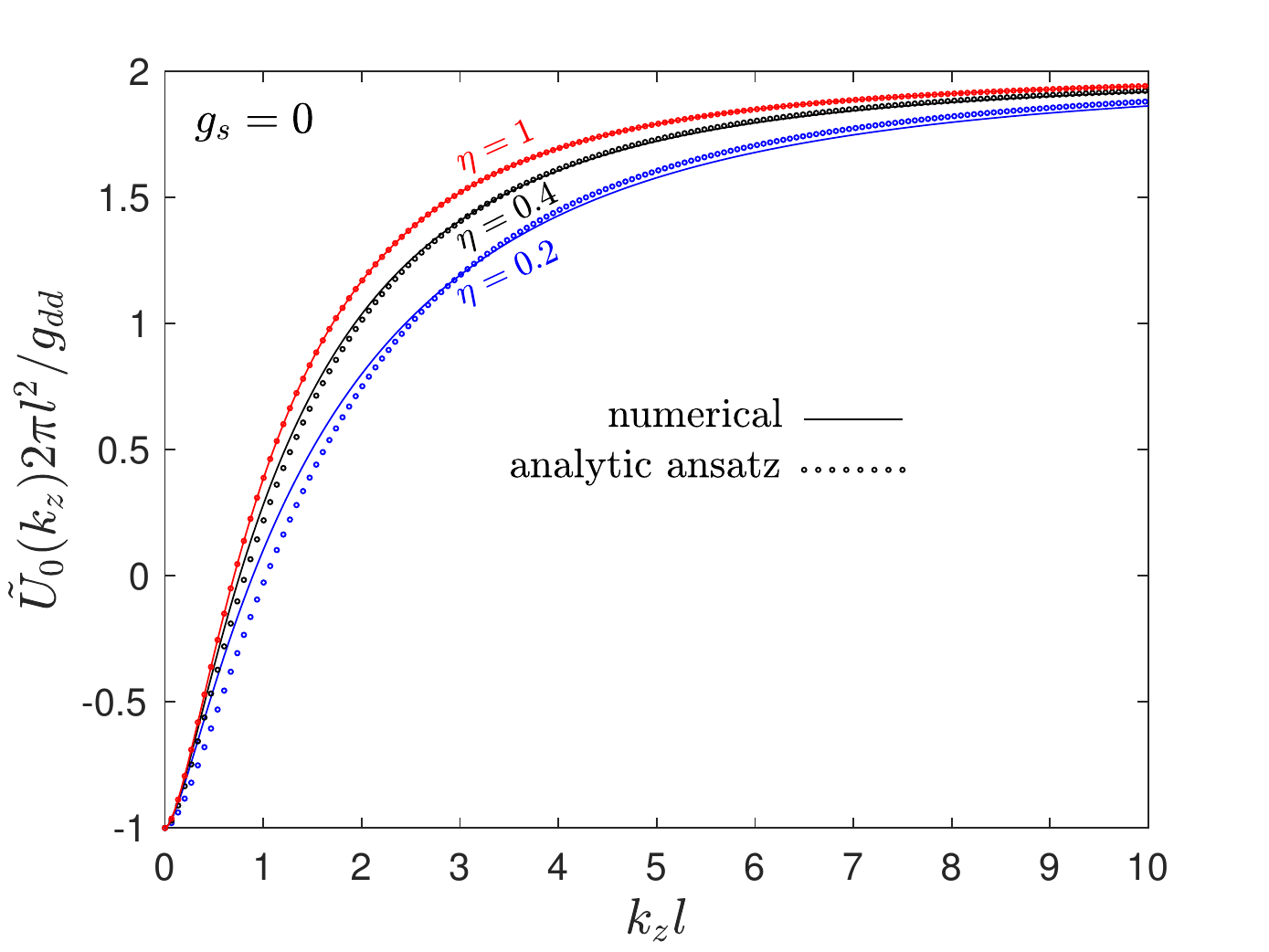}    	
      	\caption{Comparison of analytical and numerical results for the $k_z$-space interaction for the variational Gaussian function and $\alpha=0$. Results are shown for several values of $\eta$ and $g_s=0$. Since the results for $\eta$ and $1/\eta$ are identical  we only show cases with $\eta\le1$. The analytic result is given by Eqs.~(\ref{Uk}) and (\ref{Q2}) with $q_0(\eta)=(\frac{2\eta}{1+\eta^2})^{4/5}$. The numerical result is obtained by evaluating   $\tilde{U}_{0}^\text{num}(k_z)=\int \frac{d\bk_\brho}{(2\pi)^2}\tilde{U}(\mathbf{k})\left|\mathcal{F}_\brho\{(\chi^\text{var})^2\}\right|^2$.  
	}
      	\label{GVarcomparison}
      \end{figure}

 The variational theory also furnishes a description of the excitations under the same shape approximation \cite{Baillie2015b,Blakie2020a}, i.e.~by setting $u_{k_zj=0}(\bm{\rho})\to \text{u}_{k_z}\chi^\text{var}(\bm{\rho})$ and  $v_{k_zj=0}(\bm{\rho})\to \text{v}_{k_z}\chi^\text{var}(\bm{\rho})$. This approximation neglects excited transverse modes and thus captures a single excitation band.  Under this approximation the BdG excitation energy becomes
\begin{align}
E_{k_z}^\text{var}= \sqrt{\epsilon_{k_z}[\epsilon_{k_z}+2n\tilde{U}_\alpha(k_z)]} ,\label{BdGUniform}
\end{align} 
where $\text{u}_{k_z}^{ }=\cosh\vartheta_{k_z}$, $\text{v}_{k_z}^{ }=\sinh\vartheta_{k_z}$, with
$\tanh2\vartheta_{k_z}={n\tilde{U}_\alpha({k_z})}/[{\epsilon_{k_z}+n\tilde{U}_\alpha({k_z})}]$. 
 Here the $k_z$-space interaction is well-approximated as  
\begin{align}
\tilde{U}_\alpha(k_z)= \tilde{U}_\alpha(0)
+ \frac{g_{dd}}{2\pi l^2}A_\alpha Q_{\alpha}^2e^{Q_{\alpha}^2}\text{Ei}(-Q_{\alpha}^2),\label{Uk}
\end{align} 
where $A_\alpha=3\left(\frac{1-\cos^2\alpha}{1+\eta}-\cos^2\alpha\right)$. 
 For $\eta=1$ this expression gives the exact result using $Q_\alpha^2=\frac{1}{2}k_z^2l^2$  \cite{Sinha2007a}. For $\eta\ne1$ there is no exact result, however an accurate approximation can be obtained by setting 
 \begin{align}
 Q_\alpha^2=\frac{1}{2}q_\alpha(\eta)k_z^2l^2,\label{Q2}
 \end{align}
  where $q_\alpha(\eta)$ is a scaling parameter. In Ref.~\cite{Blakie2020a}  the $\alpha=\pi/2$ case was examined and  $q_{\pi/2}(\eta)=\sqrt{\eta}$ was found to
   be a good approximation. Following a similar approach we have found that for $\alpha=0$ a good approximation is obtained with $q_0(\eta)=(\frac{2\eta}{1+\eta^2})^{4/5}$ [see Fig.~\ref{GVarcomparison}].
    We view this as an important result of this work that significantly extends the applicability of the variational approach to elongated dipolar BECs. In Table \ref{tab:Uklimits} we give the exact limiting values of $\tilde{U}_\alpha$, which is important in understanding the behavior of the elongated dipolar system.

      \begin{table}
\caption{\label{tab:Uklimits}Limiting behavior of $\tilde{U}_\alpha(k_z)$.} \begin{ruledtabular}
\begin{tabular}{l|ll}
& $\tilde{U}_\alpha(k_z=0)$ & $\tilde{U}_\alpha(k_z\to\infty)$ \\[5pt] 
\hline\\[-8pt] 
 $\alpha=0$ & $\frac{1}{2\pi l^2}(g_s-g_{dd})$ &  $\frac{1}{2\pi l^2}(g_s+g_{dd})$ \\[5pt] 
 
$\alpha=\frac{\pi}{2}$ & $\frac{1}{2\pi l^2}(g_s+g_{dd}\frac{2-\eta}{1+\eta})$ &  $\frac{1}{2\pi l^2}(g_s-g_{dd})$ \\[5pt] 
 \end{tabular}
\end{ruledtabular}
\end{table}

The slope of the variational result (\ref{BdGUniform}) gives the speed of sound as
\begin{align}
c_\alpha^\text{var,0}=\sqrt{n\tilde{U}_\alpha(0)/m}.\label{c0}
\end{align}
In the beyond-quasi-1D regime the same shape approximation for the excitations tends to result in a higher speed of sound than that obtained from the full BdG calculations, and Eq.~(\ref{c0}) becomes inaccurate. 
To obtain a better estimate of the speed of sound from the variational theory we use the thermodynamic expression $c=\sqrt{n(\partial\mu/\partial n)/m}$ (also see \cite{Salasnich2004a}), which allows us to obtain an expression for the speed of sound directly from derivatives of energy functional (\ref{Eq:Eu}). This gives
\begin{align}
c_\alpha^\text{var}=\sqrt{\frac{n\tilde{U}_\alpha(0)}{m}+\frac{\partial \tilde{U}_\alpha(0)}{\partial n}\frac{n^2}{2m}},\label{varspeedofsound}
\end{align}
where the $n$-dependence of $\tilde{U}_\alpha(0)$ arises because $\eta$ and $l$ depend on $n$ [see Eq.~(\ref{U0})] (i.e.~revealing the beyond-quasi-1D behavior of the system).

 The static structure factor obtained using the variational description of the excitations is
\begin{align}
       S^\text{var}(k_z)  =\frac{\epsilon_{k_z}}{E_{k_z}^\text{var}}\left[1+2f(E_{k_z}^\text{var})\right].\label{fr}
\end{align}
 
\section{System properties}\label{sec:cases}

      \begin{table}
\caption{\label{tab:cases}Interaction parameters and features of the three cases considered in this section.} \begin{ruledtabular}
\begin{tabular}{l|lll}
& Sec.~\ref{case1} & Sec.~\ref{case2} & Sec.~\ref{case3} \\
regime & \multicolumn{2}{c}{Contact dominated}  & \multicolumn{1}{c}{DDI dominated} \\
$\bm{\mu}_m$ direction & $\alpha=0$ & $\alpha=\frac{\pi}{2}$ & $\alpha=\frac{\pi}{2}$ \\
$a_{dd}$: & $ 130.8a_0$ & $ 130.8a_0$ &$ 130.8a_0$ \\
$a_s$: & $ 140a_0$ & $ 140a_0$ & $ 115a_0$ \\
feature & \multicolumn{1}{l}{anti-roton} & $\quad$--  & \multicolumn{1}{l}{roton}\\
\end{tabular}
\end{ruledtabular}
\end{table}

      We begin by examining the basic properties of the ground states and their spectra for the two dipole orientations. This allows us to quantify the regimes where both orientations are stable, note important features in the excitation spectra and structure factors, and make some comparisons between the full meanfield theory and the  variational theory.
To illustrate the system behavior  in this section we consider a $^{164}$Dy condensate with $n=2.5\times10^3/\mu$m in an isotropic transverse trap of frequency $\omega_{x,y}/2\pi=150$Hz. 
 This regime (i.e~condensate density and transverse confinement frequency) is similar to that used in recent experiments with dipolar quantum gases in cigar shaped potentials \cite{Chomaz2018a,Petter2019a,Tanzi2019a,Bottcher2019a,Chomaz2019a}. 
 
 We consider  three different cases for the interaction parameters, which are given in  Table \ref{tab:cases}. In all cases $ {ng_s}/{2\pi l_\text{ho}^2\sim30\epsilon_\text{ho}}$ and ${ng_{dd}}/{2\pi l_\text{ho}^2\sim30\epsilon_\text{ho}}$, where $l_\text{ho}=\sqrt{\hbar/m\omega_{x,y}}$ and $\epsilon_\text{ho}=\hbar\omega_{x,y}$ are the transverse confinement length and energy scale, respectively. Importantly, this means the system is well-beyond the quasi-1D regime\footnote{The quasi-1D regime requires both ${ng_{s}}/{2\pi l_\text{ho}^2}$ and ${ng_{dd}}/{2\pi l_\text{ho}^2}$ to be smaller than $\epsilon_\text{ho}$.} necessitating the variational description as a minimal model. 
While the values of $a_s$ and $a_{dd}$ are similar (i.e.~$\sim10^2a_0$) for these cases,   the difference $a_s-a_{dd}$ being positive (i.e.~contact interaction dominated) or negative  (i.e.~DDI dominated) is important.  For example, we do not consider a DDI dominated case for $\alpha=0$ since it is mechanically unstable in this regime. On the other hand,  the $\alpha=\frac{\pi}{2}$ system is meta-state in the DDI dominated regime and can develop a roton excitation .

 \begin{figure}[tbh!]
      	\centering
      	\includegraphics[width=3.3in]{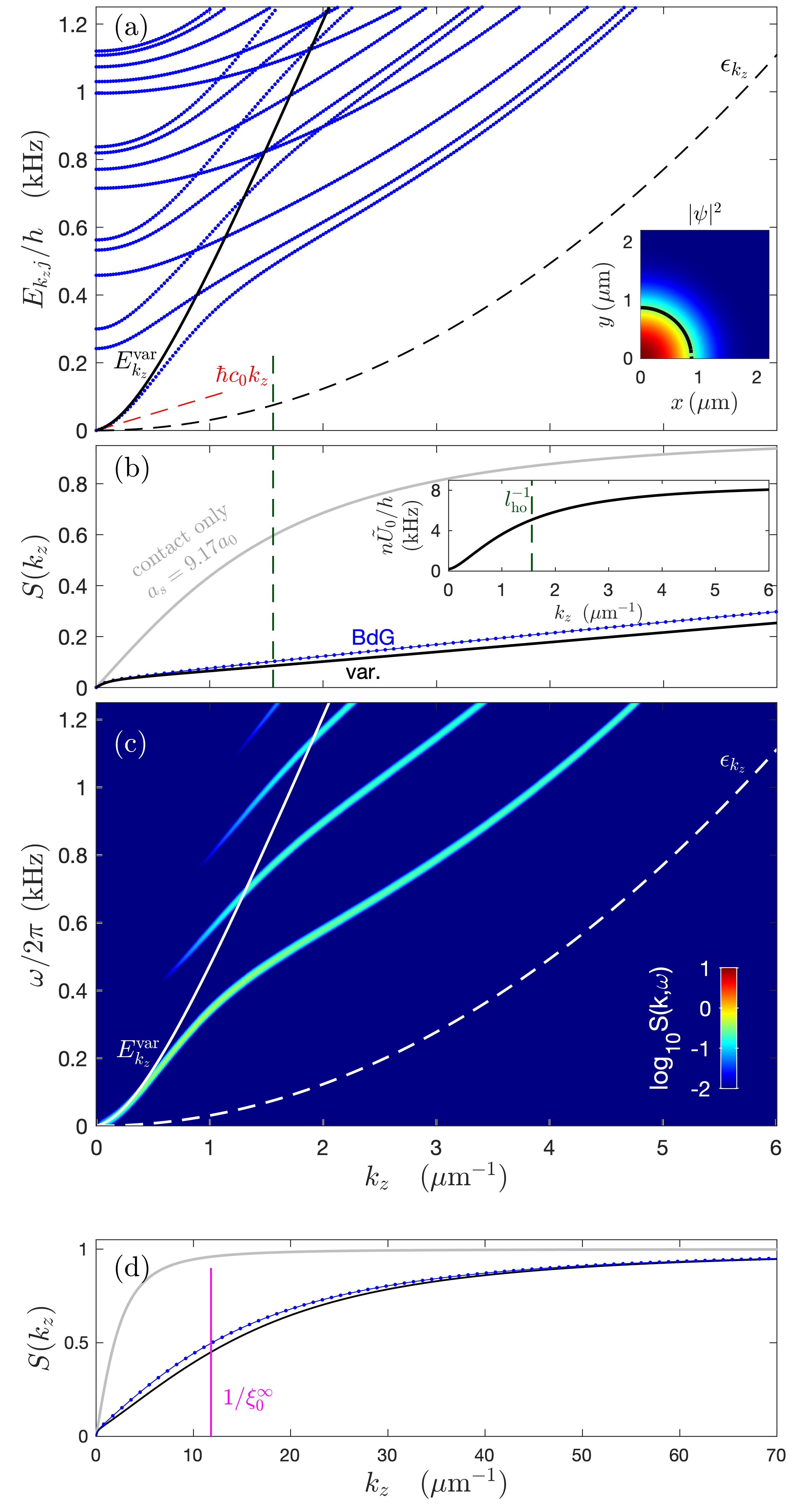}    	
      	\caption{Properties of excitations for a $^{164}$Dy elongated dipolar system with $\alpha=0$, $a_s=140a_0$ and  $a_{dd}=130.8a_0$.
	(a) The BdG excitation energy $E_{k_zj}$ for even parity  modes  (blue dots),  variational  $E_{k_z}^\text{var}$ (solid black line) and   free-particle  $\epsilon_{k_z}$ (black dashed line) dispersion relations. The red dashed line indicates the linear phonon spectrum. The inset shows  $|\chi|^2$ with the black dotted line indicating where the $|\chi^\text{var}|^2$ is 1/e of its peak value. 
(b) $S(k_z)$ at $T=0$  obtained from BdG and variational theories. (Inset) $k_z$-space interaction (\ref{Uk}). (c)  $S(k_z,\omega)$ obtained from BdG results. (d)  $S(k_z)$  as in (b), but over larger $k_z$ range. In (b) and (d) the static structure factor for  $a_{dd}=0$ and $a_{s}=9.17a_0$ (grey line) is also shown. 
	Results are for  isotropic transverse confinement $\omega_{x,y}/2\pi=150$Hz, and $n=2.5\times10^3/\mu$m. The $\delta$-functions in the definition of $S(k_z,\omega)$ are frequency broadened by Gaussian of width $50$Hz.}
      	\label{spectra0}
      \end{figure}

\subsection{Contact dominated case with $\alpha=0$ (anti-roton)}\label{case1}
The spectrum for the $\alpha=0$ case of Table \ref{tab:cases}  is given in Fig.~\ref{spectra0}(a), with the condensate density shown in the inset.  The lowest band of excitations is seen to have a small region over which it increases linearly with $k_z$ near $k_z=0$, revealing the long wavelength phonon behavior. In general the lowest excitation band is well above the free particle result $\epsilon_{k_z}$, demonstrating the importance of interactions.

The variational solution has $l=0.87\mu$m and $\eta=1$, and is in good qualitative agreement with the GPE solution. The variational spectrum   $E_{k_z}^\text{var}$ (\ref{BdGUniform}) agrees well with the lowest BdG band for $k_z\lesssim 0.5\mu$m$^{-1}$. For  $k_z\gtrsim 0.5\mu$m$^{-1}$  the variational spectrum departs from  the lowest BdG band, and is seen to ascend crossing many excited bands. 

The ground state and excitations can be qualitatively understood from $\tilde{U}_0(k_z)$ [see Eq.~(\ref{Uk}) and Table \ref{tab:Uklimits}], which characterizes the $k_z$-dependence of the interaction, and is shown in the inset to Fig.~\ref{spectra0}(b). We see that  
$\tilde{U}_0(k_z)$ monotonically increases with increasing $k_z$ and saturates at large $k_z$.  The $k_z$-dependence arises from the transverse confinement  and  $l_\text{ho}$ gives a characteristic length scale dividing the low-$k_z$ and high-$k_z$ behavior. 
The condensate interactions are described  by $\tilde{U}_0(0)$ [i.e.~see Eq.~(\ref{Eq:Eu})]. For the $\alpha=0$ case the DDI and contact terms offset at $k_z=0$ and $\tilde{U}_0(0)$ is a small and positive, so that the condensate particles are only weakly interacting.  As a result the condensate is narrow and dense, i.e.~the variational width $l$ is only slightly greater than the transverse confinement length ($l_\text{ho}=0.64\mu$m). The non-zero-$k_z$ behavior is revealed in the excitation spectrum, which directly depends on  $\tilde{U}_0(k_z)$ [see Eq.~(\ref{BdGUniform})]. In particular, the rapid increase in $\tilde{U}_0(k_z)$ with increasing $k_z$ causes the excitation energy to be shifted much higher than the free particle result $\epsilon_{k_z}$.  The difference between the variational and lowest band of the BdG spectrum at high-$k_z$ occurs because the same-shape approximation used to derive Eq.~(\ref{BdGUniform}) fails, and the excitations take a different shape to reduce overlap with the condensate.

Despite the contrasting behavior of the full BdG and variational spectra the respective  $T=0$ static structure factors  are in good agreement [see Figs.~\ref{spectra0}(b) and (d)].  We observe that the static structure factor slowly approaches the incoherent limit $S(k_z\to\infty)=1$ for $k_z\gg 1/\xi_0^\infty$, where 
$
\xi_0^\infty\equiv {\hbar}/{\sqrt{mn\tilde{U}_0(k_z\to\infty)}},
$ 
is the healing length obtained using the high-$k_z$ limit of the interactions [see Table \ref{tab:Uklimits}]. Here $1/\xi_0^\infty\approx11.8\mu$m$^{-1}$, with subplot (d) verifying that this sets the characteristic scale for $S(k_z)$ to approach the incoherent limit. For reference we also show the static structure factor for a contact interacting BEC of scattering length $a_s\approx9.17a_0$, chosen to match the speed of sound of the system with DDIs. We see that the static structure factors for both cases agree for $k_z\to0$, but that the contact result approaches the incoherent limit much more rapidly with increasing $k_z$.

 In Fig.~\ref{spectra0}(c) we show the $T=0$ dynamic structure factor (\ref{DSF}), to reveal how much each BdG excitation band contributes to the density response. At low $k_z$ only the lowest band has significant weight, whereas with increasing $k_z$ many bands contribute.  This multi-band behavior is quite novel for a condensate, where normally the lowest band carries the majority of the weight (also compare to $S(k_z,\omega)$ for the cases we examine in Secs.~\ref{case2} and \ref{case3}), and could be revealed using Bragg spectroscopy. Qualitatively we see that $E_{k_z}^\text{var}$ crosses the excited BdG bands at a similar point to when they begin contributing significant weight to the structure factor. We also note that the particular bands that are excited correspond to those with $m_z=0$ where $m_z$ is quantum number associated with the $z$-projection of the angular momentum\footnote{For a radially symmetric trap and $\alpha=0$  the system has radially symmetry. Thus $m_z$ is a good quantum number and only $m_z=0$ modes can contribute nonzero matrix elements in Eq.~ (\ref{deltank}).}.

We refer to the $\alpha=0$ case and its peculiar behavior explored in this subsection as arising from the anti-roton effect, i.e.~from an interaction that gets more strongly repulsive with increasing $k_z$.  We choose this terminology to emphasize the contrast to the usual roton regime in a dipolar BEC, which occurs when the interaction strength decreases with increasing $k_z$ (see Sec.~\ref{case3}).

\begin{figure}[tbh!]
      	\centering
      	\includegraphics[width=3.3in]{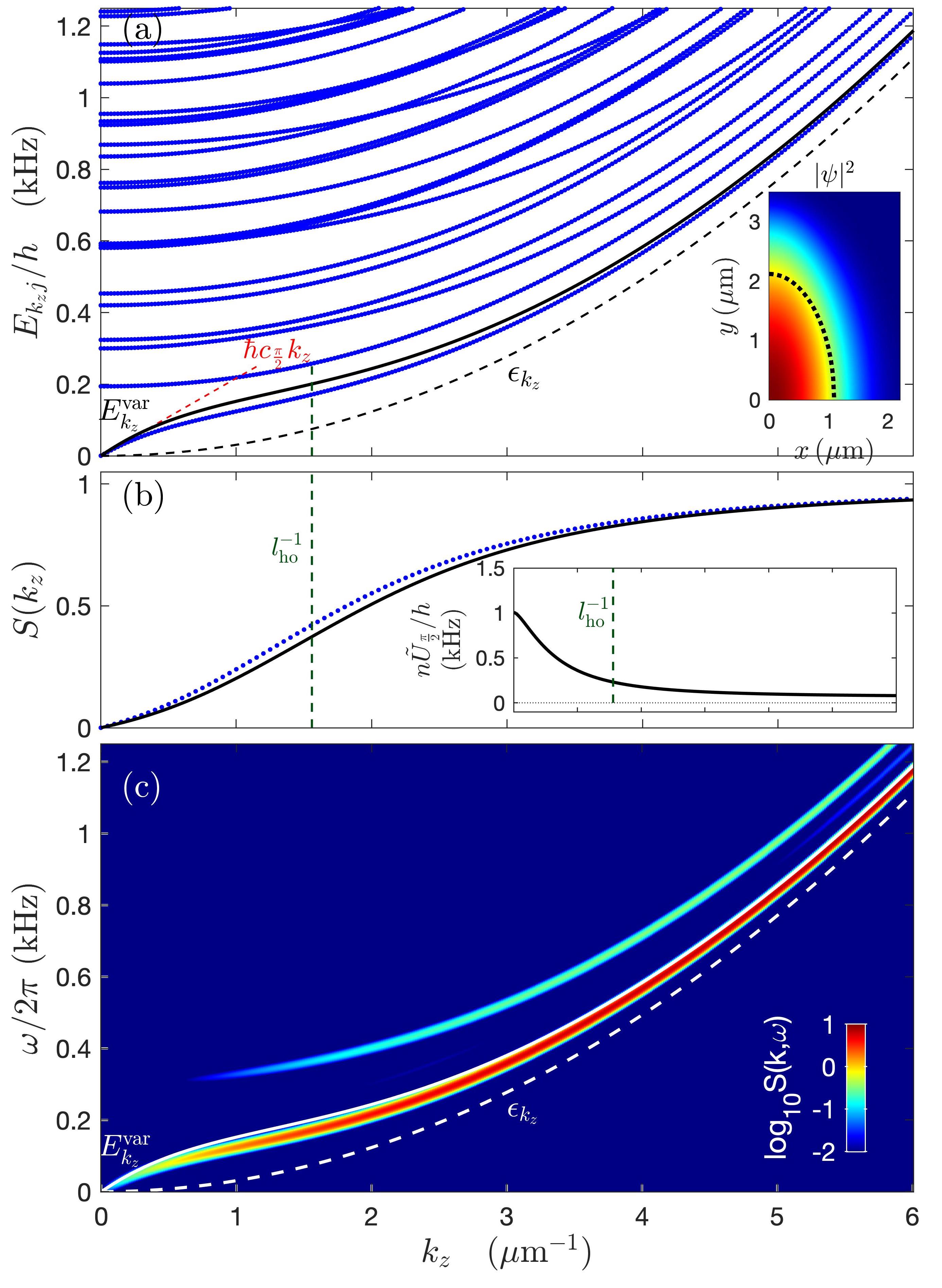}    	
      	\caption{Properties of excitations for a $^{164}$Dy elongated dipolar system with $\alpha=\pi/2$, $a_s=140a_0$ and  $a_{dd}=130.8a_0$.
	(a) The BdG excitation energy $E_{k_zj}$ for even parity modes (blue dots), with the variational  $E_{k_z}^\text{var}$ (solid black line) and  free-particle  $\epsilon_{k_z}$ (black dashed line) dispersion relations shown. The red dashed line indicates the linear phonon spectrum. The inset shows  $|\chi|^2$ with the black dotted line indicating where the $|\chi^\text{var}|^2$ is 1/e of its peak value. 
(b) Zero temperature static structure factor obtained from BdG and variational theories. (Inset) Variational $k_z$-space interaction. (c) Dynamic structure factor obtained from BdG results.  Other parameters as in Fig.~\ref{spectra0}.}
      	\label{spectrapi2A}
      \end{figure}

\subsection{Contact dominated case with $\alpha=\pi/2$ }\label{case2}
The spectrum for the $\alpha=\pi/2$ and $a_s=140a_0$ case of Table \ref{tab:cases} is given in Fig.~\ref{spectrapi2A}(a), with the condensate density shown in the inset. 
Here the condensate is anisotropic in the transverse plane arising from magnetostrictive effects of the dipoles being polarized along $y$. The variational solution has $l=1.5\mu$m and $\eta=1.9$, and is qualitatively similar to the GPE solution. The lowest BdG band of excitations is in good agreement with the variational result over the full $k_z$ range. Also the shift of the lowest band from the free particle result at high-$k_z$ is smaller than for the $\alpha=0$ case.

The ground state and excitations can be qualitatively understood from $\tilde{U}_{\pi/2}(k_z)$ shown in the inset to Fig.~\ref{spectrapi2A}(b). For this orientation of dipoles the interaction has a maximum at $k_z=0$, where the contact interaction and DDI contributions add together. This causes the condensate to experience a strong repulsive interaction and it broadens to lower its density in response. The energy is also reduced by the system distorting in the transverse plane, i.e.~$\eta$ increases to reduce $\tilde{U}_{\pi/2}(0)$ [see Table \ref{tab:Uklimits}]. 
As $k_z$ increases $\tilde{U}_{\pi/2}(k_z)$ monotonically decreases, saturating to a small repulsive interaction (since $g_{s}>g_{dd}$ for this case) at  high-$k_z$.    As in the $\alpha=0$ case $l_\text{ho}$ gives a characteristic length scale dividing the low-$k_z$ and high-$k_z$ behavior.    The weak repulsive interactions at high-$k_z$ explain the relatively small shift in the lowest BdG band from the free particle dispersion relation. The $T=0$ static structure factor Fig.~\ref{spectrapi2A}(b) shows good agreement between the variational and BdG results, with the dynamic structure factor [Fig.~\ref{spectrapi2A}(c)] showing that most of the weight comes from the lowest band.

\begin{figure}[tbh!]
      	\centering
      	\includegraphics[width=3.3in]{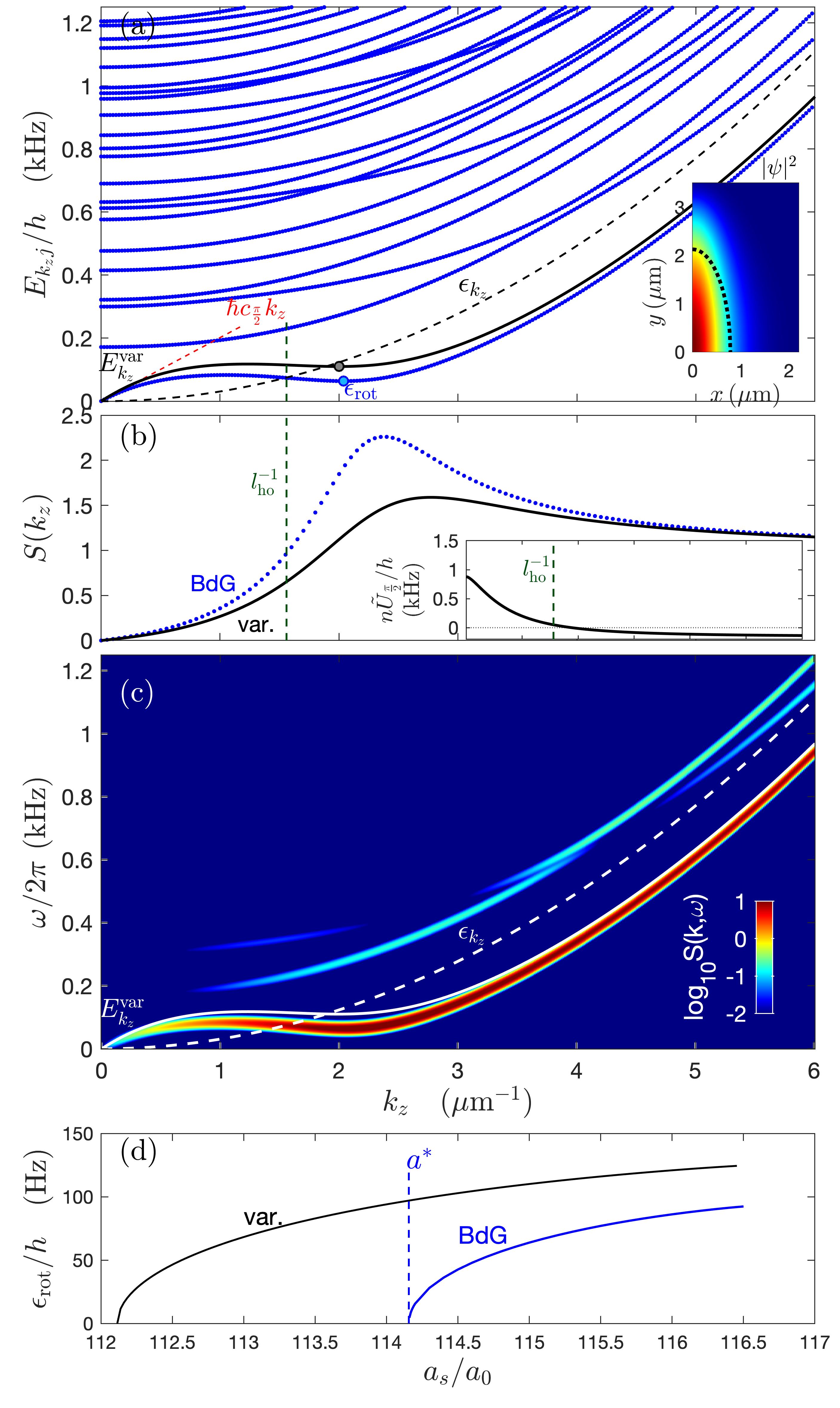}    	
      	\caption{Properties of excitations for a $^{164}$Dy elongated dipolar system with $\alpha=\pi/2$ and $a_s=115a_0$.
	(a) The BdG excitation energy $E_{k_zj}$ for modes that are of even parity in $x$ and $y$ (blue dots), with the variational  $E_{k_z}^\text{var}$ (solid black line) and the free-particle dispersion $\epsilon_{k_z}$ (black dashed line) dispersion relations shown. The red dashed line indicates the linear phonon spectrum using the variational speed of sound.  The local minimum of the dispersion relations  identifies the roton excitation (filled circles). The inset shows  $|\chi|^2$ with the black dotted line indicating where the $|\chi^\text{var}|^2$ is 1/e of its peak value. (b) Zero temperature static structure factor obtained from BdG and variational theories. (Inset) Variational $k_z$-space interaction. (c) Dynamic structure factor obtained from BdG results. (d) The roton energy from the variational and BdG calculation as a function of $a_s$, shown from when it appears until it goes to zero energy.  Other parameters as in Fig.~\ref{spectra0}.}
      	\label{spectrapi2B}
      \end{figure}

\subsection{DDI  dominated case with $\alpha=\pi/2$ (roton)} \label{case3}
The DDI dominated case with $\alpha=\pi/2$  and $a_s=115a_0$ shares many features with the case considered in the last subsection, and here we mainly focus on the new aspects. 

For this case  the variational solution has $l=1.3\mu$m and $\eta=2.7$. Being in the DDI dominated regime, i.e.~$a_s<a_{dd}$,  the  effective interaction $\tilde{U}_{\pi/2}(k_z)$ becomes negative at high $k_z$ [see inset to Fig.~\ref{spectrapi2B}(b), and Table \ref{tab:Uklimits}]. This attractive interaction causes a roton excitation to form, identified as a local minimum in the dispersion relation.  We have labeled the energy of this minimum as $\epsilon_\text{rot}$ in  Fig.~\ref{spectrapi2B}(a) and denote the wavevector where it occurs as  $k_\mathrm{rot}$.  Because $l_\text{ho}$ gives a characteristic length scale where $\tilde{U}_{\pi/2}$ switches over to being attractive, we have $k_\text{rot}\sim1/l_\text{ho}$.  The $T=0$ static structure factor Fig.~\ref{spectrapi2B}(b) now exhibits non-monotonic behavior and has a peak at intermediate $k_z$ values  near $k_\text{rot}$.  The qualitative agreement between the variational and BdG results for $S(k_z)$ is less good in this regime because the magnitude of the peak is sensitive to the energy of the roton, which in turn is   sensitive to the precise details of the system.  In Fig.~\ref{spectrapi2B}(d)  we show how the roton energy varies with $a_s$. It first emerges around $a_s=116.5a_0$ and then decreases with decreasing $a_s$. The BdG results show that the roton goes to zero energy at a critical $a_s$ value of $a^*\approx114.2a_0$ at which point the system becomes unstable. The variational result predicts this instability to be at the lower value of $112.1a_0$.

\begin{figure}[tbh!]
      	\centering
      	\includegraphics[width=3.3in]{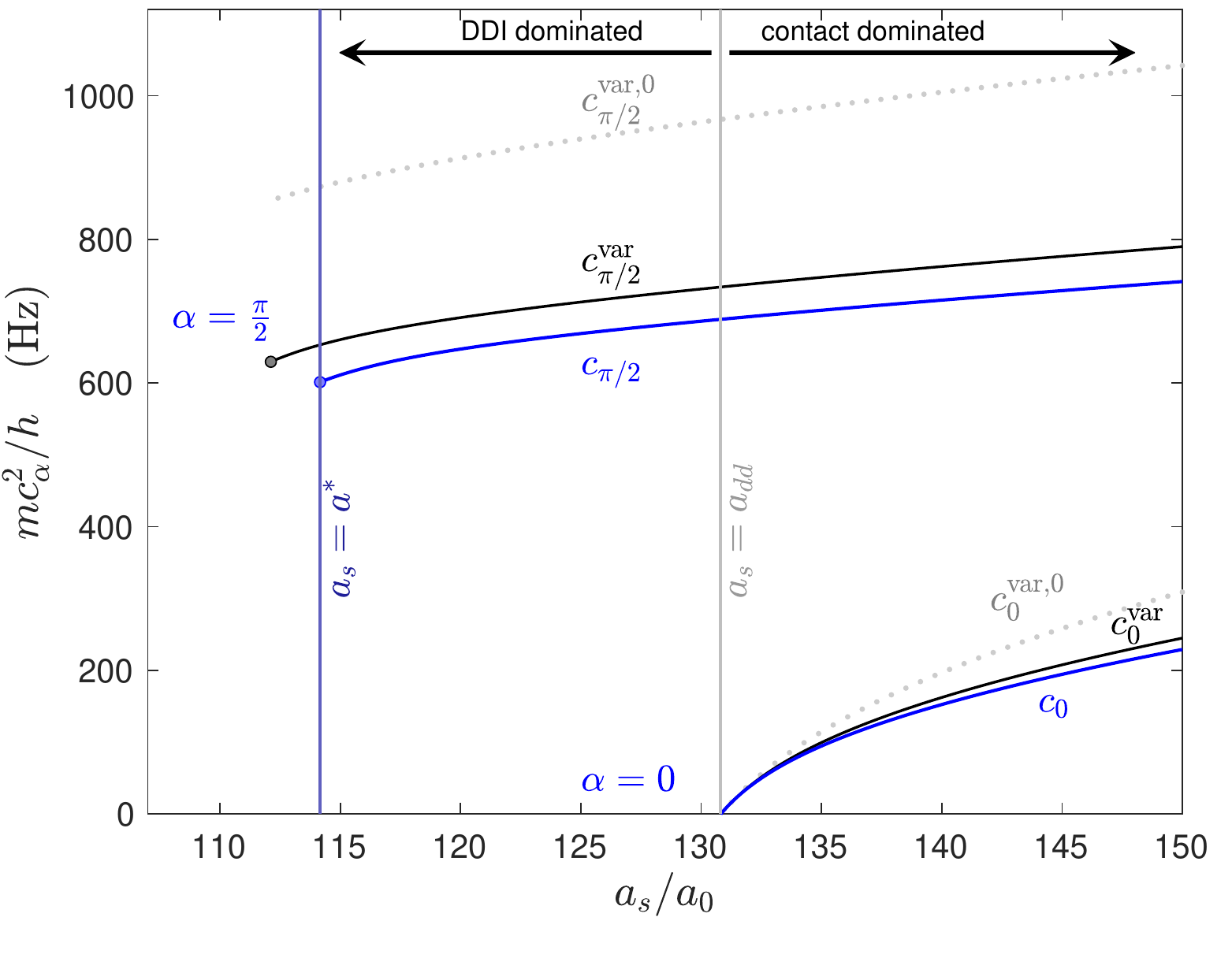}    	
      	\caption{Speed of sound and stability. The speed of sound versus $a_s$ for an elongated  dipolar system at $\alpha=0$ (lower curves) and $\alpha=\pi/2$ (upper curves). We show both the BdG and variational results [see Eqs.~(\ref{cBdG}), (\ref{varspeedofsound}) and  (\ref{c0})]. The left hand termination point of the curves marks where the system is unstable.	Other parameters as in Fig.~\ref{spectra0}.}
      	\label{speedofsound}
      \end{figure}
      
\subsection{Speed of sound and stability}
In Fig.~\ref{speedofsound} we show the speed of sound over a wide parameter regime, including the cases studied in Figs.~\ref{spectra0}-\ref{spectrapi2B}. The speed of sound is determined by the $k_z\to0$ interactions of the system [see Table \ref{tab:Uklimits}], and most strikingly we see in these results that even though the density is the same, the speed of sound for $\alpha=\pi/2$ is much higher than the $\alpha=0$ case. We also see that the $\alpha=0$ system becomes unstable when $a_s$ reduces to be equal or less than $a_{dd}$. At this point the speed of sound goes to zero (i.e.~the compressibility diverges) and the gas mechanically collapses. For the $\alpha=\pi/2$ case, the unstable point  instead arises from the roton excitation goes soft, occurring when $a_s\approx114.2\,a_0$ [see Fig.~\ref{spectrapi2B}(d)].

We also compare the variational and BdG predictions for the speed of sound in Fig.~\ref{speedofsound}. This shows that the variational formalism provides an accurate estimate of the speed of sound, and that the corrected form presented in Eq.~(\ref{varspeedofsound})  significantly improves upon the bare variational result of Eq.~(\ref{c0}).

\section{Number fluctuation results}\label{results} 
In Fig.~\ref{mainflucts} we show the relative number fluctuations, $\Delta N_L^2/N_L$, as a function of cell size for the three cases considered in Sec.~\ref{sec:cases}. For comparison we also include the results of a similar calculation for a contact interacting condensate with $a_s=140a_0$. We show results for zero [Fig.~\ref{mainflucts} (a)] and non-zero  [Fig.~\ref{mainflucts}(b),(c)] temperatures. Our full numerical results are obtained by constructing the static structure factor from the BdG solution on a dense $k_z$ grid and then numerically integrating Eq.~(\ref{var}) for each cell size $L$.

\begin{figure}[tbh!]
      	\centering
      	\includegraphics[width=3.3in]{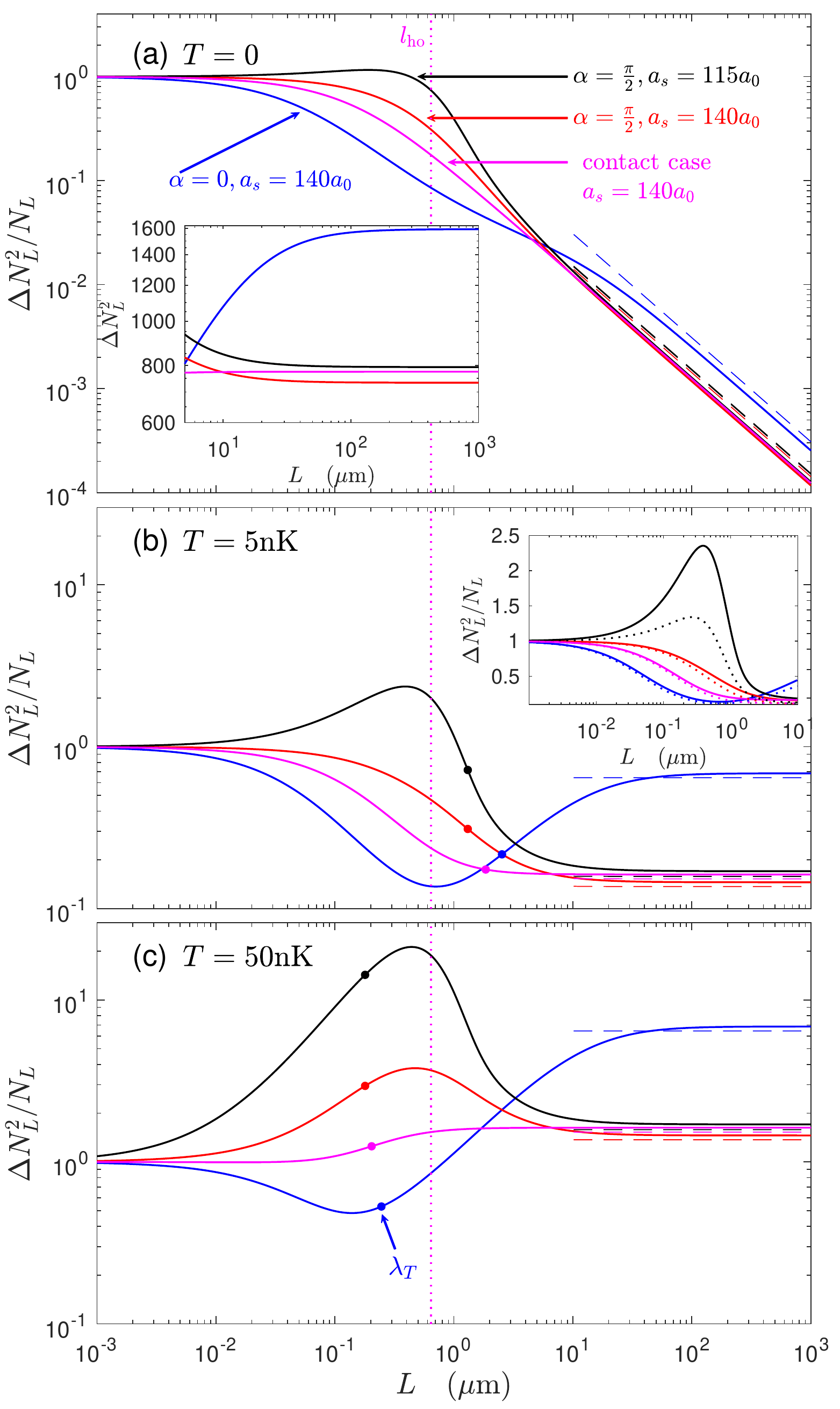}    	
      	\caption{Relative number fluctuations of an elongated dipolar condensate versus linear cell size for (a)  $T=0$, (b) $T=5\,$nK, and (c) $T=50\,$nK.  Solid lines are calculated using the GPE ground state and BdG excitations for the three cases considered in Sec.~\ref{sec:cases} and a contact interactions only case with  $a_s=140a_0$. The dashed lines are approximate large-cell results based on the variational speed of sound (see text)  plotted for $L>10\mu$m, and the filled circles in (b) and (c) indicate  $\lambda_T$.   In all subplots the transverse confinement length is shown for reference.
	The inset of (a) shows the (unscaled) number fluctuations calculated at $T=0$ using  the GPE ground state and BdG excitations for large cells. The inset of (b) shows the full results (solid lines) compared to  the numerically integrated variational result (dotted) (see text). }
      	\label{mainflucts}
      \end{figure}

\subsection{Small and large cell behavior}
Independent of temperature, for small cell sizes the fluctuations approach the Poissonian limit, where $\Delta N^2_L$ is equal to the mean number of atoms in the cell, $N_L$. This occurs because the particles are uncorrelated on small length scales.

 In the large cell limit the behavior is strongly dependent on the temperature.   At zero temperature the relative fluctuations decrease  as $\sim L^{-1}$ with increasing cell size.    Since $N_L\propto L$, this means that $\Delta N^2_L\to\text{const}$, as revealed in the inset to Fig.~\ref{mainflucts}(a). 
 At non-zero temperatures the relative fluctuations plateau to a constant value, i.e.~realizing the thermodynamic behavior of Eq.~(\ref{thlimit}). This plateau is strongly dependent on the dipole orientation $\alpha$, as the thermodynamic limit is sensitive to the speed of sound (cf.~Fig.~\ref{speedofsound}) and thus the $k_z\to0$ behavior of the interactions (cf.~Table \ref{tab:Uklimits}). Notably, for $\alpha=0$ where the DDIs significantly reduce the speed of sound, we observe stronger fluctuations in the large cell limit. For the $\alpha=\pi/2$ cases, the fluctuations in the thermodynamic limit are similar to the contact interacting case. The speed of sound for dipolar systems in this case is sensitive to magnetostriction   (e.g.~the role of $\eta$ in $\tilde{U}_{\pi/2}(0)$, see Table \ref{tab:Uklimits}).

For comparison to our full numerical results in the large cell limit we can develop a simple approximation based on the variational speed of sound from Eq.~(\ref{varspeedofsound}). Setting  $E_{k_z}^\text{var}=\hbar c_\alpha^\text{var}k_z$ ($k_z\ll1/l_\text{ho}$) the leading order behavior of the static structure factor is
 \begin{align}
 S(k_z\to0)&=\frac{\hbar |k_z|}{2mc_\alpha^\text{var}},\qquad T=0,\\
 S(k_z\to0)&=\frac{k_BT}{m(c_\alpha^\text{var})^2},\qquad T>0.
 \end{align}
 We then make the approximation 
 \begin{align}
 \Delta N_L^2\sim N_LS(1/L),\label{DN2approx}
 \end{align}
  which assumes that the fluctuations are dominated by excitations with a wavelength comparable to the cell size (e.g.~see \cite{Klawunn2011a}).
  This  yields the dashed lines in Fig.~\ref{mainflucts}. We note that the finite temperature result using this expansion is just the thermodynamic limit (with the variational speed of sound), while the zero temperature result gives the $L^{-1}$ scaling found in the numerical results. The slight shift of these results from the full numerical solution arises from the difference in the variational speed of sound from the BdG result.
  We also mention that the thermodynamic limit as $T\to0$ is non-trivial since the low temperature thermal wavelength\footnote{In general we  define the thermal wavelength as $\lambda_T=1/k_z$, where $k_z$ satisfies $E_{k_z,j=0} = k_BT$. E.g., for low temperatures $E_{k_z,j=0}\approx \hbar c_\alpha k_z$ and we have $\lambda_T=\hbar c_\alpha/k_BT$.}   $\lambda_T=\hbar c_\alpha/k_BT$ diverges, while the thermodynamic result  only holds for $L\gg \lambda_T$.

\begin{figure}[tbh!]
      	\centering
      	\includegraphics[width=3.3in]{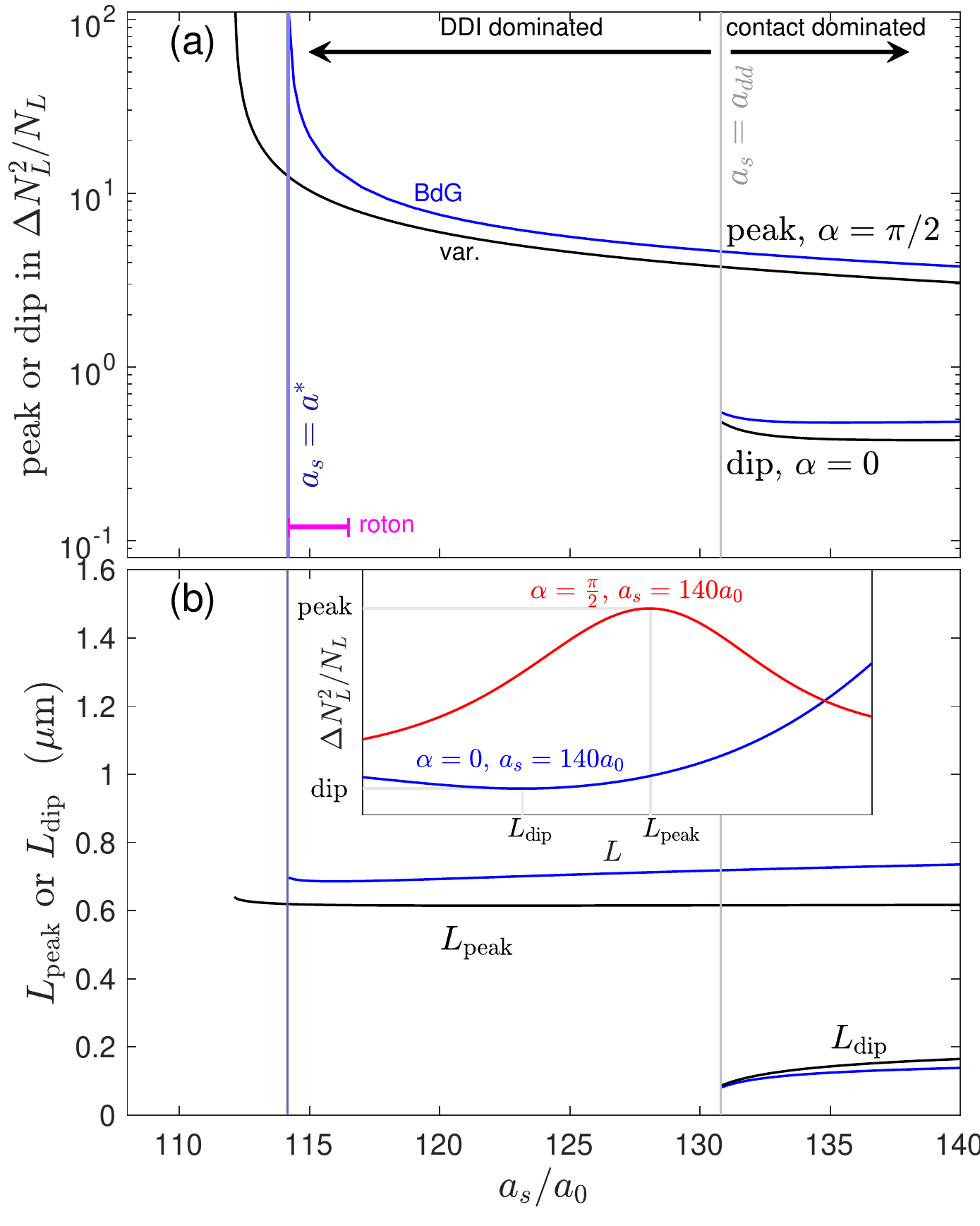}    	
      	\caption{Peak and dip features in the relative number fluctuations. (a) The maximum (peak) and minimum (dip) value of $\Delta N_L^2/N_L$  for $\alpha=\pi/2$ and $\alpha=0$, respectively. Results shown for   $T=50$nK from  BdG  (blue lines) are variational  (black lines) theories. The small magenta bar indicates the $a_s$ range over which the roton occurs in the spectrum of the $\alpha=\pi/2$ system.
	(b) The cell size $L_\text{peak}$ and $L_\text{dip}$ where the peak and dip occur, respectively, corresponding to the results in (a). The inset to (b) uses the $a_s=140a_0$ results from Fig.~\ref{mainflucts}(c) to illustrate the peak and dip in the relative number fluctuations for the $\alpha=\pi/2$ and $\alpha=0$ cases, respectively.  }
      	\label{peakflucts}
      \end{figure}

\subsection{Intermediate cell sizes: peak and dip features}
      An interesting feature of the results in Fig.~\ref{mainflucts} is the non-monotonic behavior of the relative number fluctuations with cell size. This feature occurs in many of the dipolar results, but is not observed in the contact interacting case.
  The inset to Fig.~\ref{mainflucts}(b) compares the full BdG results to the fluctuations evaluated from Eq.~(\ref{var}) using the variational static structure factor. This shows that for most cases the two approaches are qualitatively in good agreement over the full range of cell sizes considered. The peak for the $a_s=115a_0$ result is in poorest agreement. This is because the precise details of the roton are important, and the variational theory predicts a higher  roton energy in this regime [see~Fig.~\ref{spectrapi2B}(d)]. However, we conclude that in general the variational theory we have developed provides a qualitatively good description.
  
In the $\alpha=0$ elongated dipolar condensate we observe a suppression in fluctuations for moderate cell sizes. This is related to the suppression in density fluctuations at moderate $k_z$ that we previously noted in the static structure factor  (see Fig.~\ref{spectra0}). For all temperatures considered and at  intermediate cell sizes (i.e.~$L\sim l_\text{ho}$) the $\alpha=0$ system has the lowest  relative number fluctuations compared to the other cases in Fig.~\ref{mainflucts}. This feature becomes a local minimum  (i.e.~dip) when the temperature is in the range $0.2\,$nK$<T<300\,$nK. The lower temperature limit is when the thermodynamic plateau is first high enough to create a minimum. The upper temperature limit is when thermal excitation of modes with $k_z\sim 1/l_\text{ho}$ is strong enough to overcome the suppression\footnote{This behavior is well captured by the variational structure factor (\ref{fr}), which approximately relates to the number fluctuations through Eq.~(\ref{DN2approx}). Here the intrinsic suppression is described by the $T=0$ static structure factor $S_0\equiv\epsilon_{k_z}/E_{k_z}^\text{var}$, while the thermal excitation is described by $2f+1$.}. Most experiments operate at temperatures well within this temperature range and the dip feature should be observable.

In the $\alpha=\pi/2$ cases we observe enhancement of fluctuations at moderate cell sizes.
This relates to the enhancement in density fluctuations at moderate $k_z$ values in the static structure factor noted earlier  (see Figs.~\ref{spectrapi2A} and \ref{spectrapi2B}). 
For all temperatures considered and at intermediate cell sizes  the $\alpha=\pi/2$ systems have larger relative fluctuations than the other cases in Fig.~\ref{mainflucts}.  This can emerge as a local maximum (i.e.~peak) depending on the temperature and the relative strength of the dipole-dipole and contact interactions. For the  $a_s=115a_0$ case, this peak is apparent at all temperatures, including $T=0$. This occurs because the interactions are attractive for $k_z\gtrsim1/l_\text{ho}$. For the  $a_s=140a_0$ case, where the interactions are weakly repulsive at high-$k_z$, a peak occurs when the modes with enhanced fluctuations become thermally activated. In the $T=5\,$nK results of Fig.~\ref{mainflucts}  these modes are not accessed (i.e.~$\lambda_T>l_\text{ho}$), and there is no peak. While for the $T=50$nK results these modes are activated and a peak develops.

      In Fig.~\ref{peakflucts} we explore how the peak and dip features in the relative fluctuations change with $a_s$, and show the cell size where these features occur. These results show that the peak and dip features occur at all values of $a_s$ considered where the respective system is stable.
 In Fig.~\ref{peakflucts}(a) we see the peak feature for $\alpha=\pi/2$ diverges as the roton softens [cf.~Fig.~\ref{spectrapi2B}(d)]. We note that the roton only appears in the excitation spectrum in the small range $a^*<a_s<116.5a_0$, thus a peak in the relative fluctuations does not directly relate to the roton.     
 We also compare the full numerical results for the peak and dip features to those extracted from the variational approximation and find qualitative good agreement over a wide parameter  
regime.

     \vspace{0.5cm}
      \section{Conclusions}
       \label{conclusions}
       
        In this work we have studied the properties of an elongated dipolar condensate, with a focus on excitations and local number fluctuations. We have considered the role of dipole orientation, demonstrating that the excitation properties are markedly different for the dipole being along or orthogonal to the long axis of the system. In particular the speed of sound is extremely sensitive to the dipole orientation, and this is revealed in the number fluctuations for large cells.  
   Also the dipole orientation can cause a peak or dip to emerge in the fluctuations as a function of cell size,  related to the momentum dependence of the DDIs in the confined system.
    In addition to full numerical calculations of this system using the GPE and BdG equations we present a simple variational theory. We have used this theory to calculate the spectrum, including the speed of sound, structure factors and fluctuations. Importantly, the variational theory provides a good description over a wide range of experimentally relevant parameters (e.g.~see \cite{Chomaz2018a,Petter2019a,Tanzi2019a,Bottcher2019a,Chomaz2019a}), and is much easier to calculate.
The type of measurements needed to count atom numbers in finite cells has already been demonstrated in previous experiments with a Rb condensate in an elongated trap \cite{Jacqmin2011a,Armijo2011a} and should be readily applied to current dipolar condensate experiments. Measuring the peak or dip feature in the fluctuations will require having \textit{in situ} imaging resolution $\sim l_\text{ho}$, which is possible with high numerical aperture imaging systems. Our results also demonstrate that these features will be prominent for typical temperatures realized in experiments (i.e.~$\sim10^1$--$10^2\,$nK). Furthermore, if it is possible to determine system densities and interactions accurately, fluctuation measurements might be a useful form of thermometry.

  In this work we have neglected the effects of beyond meanfield corrections. For instance, these can give rise to a supersolid state for $\alpha=\pi/2$ and $a_s<a^*$ or to a macro-droplet for  $\alpha=0$ and $a_s<a_{dd}$. We expect that these corrections will have only a small effect for the regime we consider, but they will cause a slight shift in the location of the roton softening, which is very sensitive to system parameters  (e.g.~see \cite{Blakie2020a}). 
A general discussion of beyond meanfield effects on fluctuation measurements would be an interesting extension for future work, and certainly could be an avenue to get better insight into the liquid and solid-like transitions seen in dipolar gases (e.g.~see \cite{Bottcher2019b,Baillie2017a,Natale2019a,Hertkorn2019a}).

      \section{Acknowledgements}
We acknowledge support from the Marsden Fund of the Royal Society of New Zealand, and valuable discussions with L.~Chomaz.
         
%
       
      \end{document}